\documentclass[pra,twocolumn,floatfix,superscriptaddress]{revtex4-1}
\usepackage{dcolumn,amsmath}
\usepackage{graphicx}
\usepackage{bm}
\usepackage{amssymb}
\usepackage{multirow}

\usepackage{amsfonts}
\usepackage{amsmath}
\usepackage{cancel}
\usepackage{xcolor}
\begin{document}
\title{Sensitivity  of the YbOH molecule to $\mathcal{P}$,$\mathcal{T}$-odd effects in the external electric field}
\author {Alexander Petrov}\email{petrov\_an@pnpi.nrcki.ru}
\author{Anna Zakharova}  
%
\affiliation{Petersburg Nuclear Physics Institute named by B.P. Konstantinov of National Research Centre
"Kurchatov Institute", Gatchina, 1, mkr. Orlova roshcha, 188300, Russia}
\affiliation{St. Petersburg State University, St. Petersburg, 7/9 Universitetskaya nab., 199034, Russia} 

\date{Received: date / Revised version: date}
%
\begin{abstract}{
Electron electric dipole moment (eEDM) search using lasercoolable triatomics like YbOH is one of the most sensitive probes for physics beyond the Standard Model. The eEDM-induced energy shift is proportional to 
polarization ($P$) of the molecule. 
Similarly to diatomics with $\Omega-$doubling structure it was assumed that for triatomics with $l-$doubling structure, related to the vibrational angular
momentum,
 $P$ can easily be saturated to
almost 100\% value with moderate external electric field.
We developed the method 
for calculation of properties of triatomic molecules 
and applied it to calculation of $P$  of $^{174}$YbOH in the first excited $v=1$ bending mode. Calculations showed that the most of the levels reach less than 50\% efficiency.
We showed that this fact is related to the Hund's case $b$ coupling scheme of YbOH. As coupling scheme (for $\Omega=1/2$ molecules) approaches $a$ (or $c$) case polarization increases up to 100\% value.
Results of our calculations should be used for correct extracting of eEDM value from YbOH experiment
and similar calculations are required for other molecules.

} 
\end{abstract}
\maketitle
%
Measuring the electron electric dipole moment (eEDM) is now considered as a most promising test for existence of physics beyond the Standard model \cite{Fukuyama2012,PospelovRitz2014,YamaguchiYamanaka2020,YamaguchiYamanaka2021}. 
The current limit on the electron electric dipole moment (the ACME II experiment), 
$|d_e|<1.1\times 10^{-29}$ e$\cdot$cm (90\% confidence),
was set by measuring the spin precession  
using
thorium monoxide (ThO) molecules in the metastable electronic H$^3\Delta_1$ state~\cite{ACME:18}. For successful performing of the such kind of experiments the possibility of suppressing of the systematic effects is of high importance.
 Previously it was shown  that due to existence of $\Omega$-doublet levels the experiments for searching of the $\mathcal{P}$,$\mathcal{T}$-odd effects on the ThO \cite{ACME:18,DeMille:2001,Petrov:14,Vutha:2010,Petrov:15,Petrov:17} 
or the HfF$^{+}$ \cite{Cornell:2017,Petrov:18} are very robust against a number of systematics. 

In turn cold polar molecules provide unique opportunities for further progress in search for effects of symmetry violation \cite{Isaev:16}.
In such molecules the sensitivity of the experiments can be strongly enhanced due to increased coherence time. Both the possibility of laser cooling and the existence of the close levels of the opposite parity can be realized with triatomic molecules such as the RaOH \cite{Isaev_2017}, the YbOH \cite{Kozyryev:17} etc. In this case the role of the $\Omega$-doublets used in the diatomic molecular experiments is taken over by the $l$-doublets of the excited $v=1$ bending vibrational modes \cite{Kozyryev:17,Pilgram:21}. 

Any eEDM experiment searches for an eEDM induced Stark shift
\begin{equation}
\delta E = d_e E_{\rm eff}   P,
\label{split}
\end{equation}
where $d_e$ is the value of electron electric dipole moment, $E_{\rm eff}$ is {\it effective electric field} acting on electron in the molecule, which is subject of molecular calculations
\cite{denis2019enhancement,prasannaa2019enhanced,gaul2020ab,Zakharova:21a, Zakharova:21b},
$P$ is the polarization of the molecule by the external electric field.
To extract $d_e = \delta E / (E_{\rm eff} P) $ from the measured
shift $\delta E $, one needs to know $E_{\rm eff}P$.
Therefore value of $E_{\rm eff}P$ directly influence $d_e$ value extracted from experiments \cite{Nataraj2011, Porsev2012}.

It is well known that for diatomics due to the existence of $\Omega$-doublet structure $P$ value approaches unity for small laboratory electric fields
\cite{DeMille:2001}. The same situation was expected for linear triatomic molecules \cite{Kozyryev:17}. In our letter we show that the $l-doubling$ structure,
is, in general, different from $\Omega-doubling$, and the $P$ value tends to approach half of the maximum value for molecules with Hund's case $b$ coupling scheme and 100\% value as coupling scheme approaches $a$ (or $c$) case.

  In the present letter we developed the method for computation of energy levels and different properties of linear triatomic molecules. We applied it for calculation of the sensitivity of the $^{174}$YbOH molecule to eEDM in the ground rotational $N=1$ level of first excited $v=1$ bending mode to eEDM in the external electric field.
  According to eq. (\ref{split}), these calculations should be used to determine the limit on the eEDM.
  
  Together with eEDM one always needs to consider scalar T,P-odd
electron-nuclear interaction, since its influence on the spectrum of molecules is identical to eEDM. For short, only the influence of eEDM is mentioned in the present letter.


\begin{figure}[h]
\centering
  \includegraphics[width=0.48\textwidth]{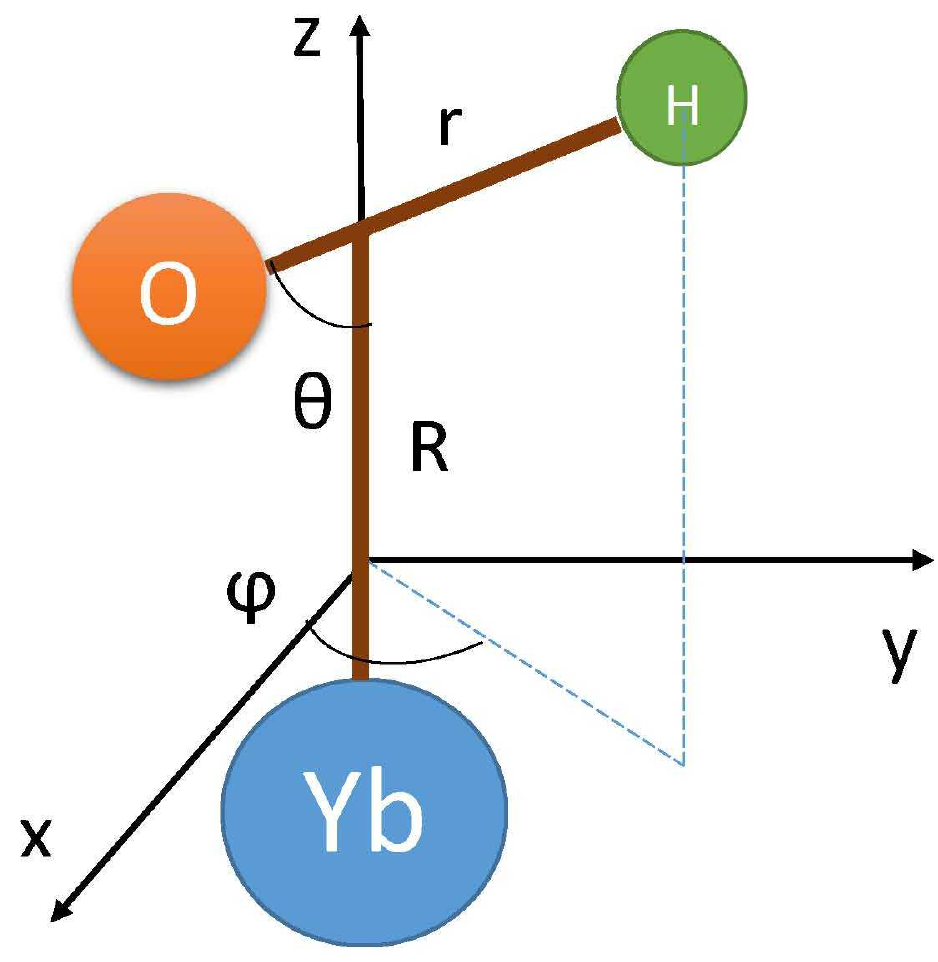}
  \caption{Orientation of the YbOH relative to the molecular frame. Orientation of the molecular frame relative to the laboratory one is given by Euler's angles (not presented in the figure) $\alpha$, $\beta$, $\gamma=0$. $R,r,\theta$ are Jacobi coordinates.}
  \label{Jacob}
\end{figure}

Consider the molecular Hamiltonian for the electronic-nuclear motion, excluding nuclear hyperfine structure, and with fixed OH ligand stretch. In the Jacobi coordinates (represented in Fig.~\ref{Jacob}) in {\it laboratory} frame it reads as
\begin{equation}
\hat{\rm H}_{\rm mol}=-\frac{\hbar^2}{2\mu}\frac{\partial^2}{\partial R^2}+\frac{\hat{\bf L}^2}{2\mu R^2}+\frac{\hat{\bf l}^2}{2\mu_{OH}r^2}+ {\rm  \hat{H}}_{\rm el},
\label{Hlabf}
\end{equation}
where $\mu$ is the reduced mass of the $Yb-OH$ system, $\mu_{OH}$ is the reduced mass of the OH ligand, $\hat{\bf L}$  is the angular momentum of the rotation of the Yb atom and OH
around their center of mass, $\hat{\bf l}$ is the angular momentum of the rotation of the OH, and ${\rm \bf \hat{H}}_{\rm el}$ is the electronic Hamiltonian.
For number of applications it is convenient to rewrite the Hamiltonian in molecular reference frame defined by Euler angles  $\alpha,\beta,\gamma$ (see e.g. Fig 20 in Ref. \cite{LL77}), with $\alpha,\beta$ corresponding to azimuthal and polar angles of axis (it will be $z$ axis of the molecular frame) going through Yb and the center of mass of OH, and $\gamma=0$. In this case the Hamiltonian (\ref{Hlabf}) is rewritten as
\begin{equation}
\hat{\rm H}_{\rm mol}=-\frac{\hbar^2}{2\mu}\frac{\partial^2}{\partial R^2}+\frac{(\hat{\bf J} -\hat{\bf J}^{e-v} )^2}{2\mu R^2}+\frac{\hat{\bf l}^2}{2\mu_{OH}r^2}+ {\rm  \hat{H}}_{\rm el},
\label{Hmolf}
\end{equation}
\textit{where $\hat{\bf J}$ is the total molecular 
less nuclear spins angular momentum}, $\hat{\bf J}^{e-v} = \hat{\bf J}^{e} + \hat{\bf J}^{v}$, $\hat{\bf J}^{e}$ - is the total electronic momentum, $\hat{\bf J}^{v}$ is the vibrational momentum,
$\hat{ J}^{v}_z = -i\hbar\frac{\partial}{\partial \varphi}$,
$\varphi$ is the angle between $xz$ plane of the molecular reference frame
and the plane of the molecule (see Fig.~\ref{Jacob}), $\hat{\bf l}^2 = \frac{1}{\sin \theta}\frac{\partial}{\partial \theta}\sin \theta \frac{\partial}{\partial \theta} + \frac{1}{\sin^2 \theta} \frac{\partial^2}{\partial \varphi^2}$, and $\theta$ is the angle between OH  and $z$ axes. The condition $\theta=0$ corresponds to the linear configuration where O atom is between Yb and H ones.

For the purpose of the present letter, we also include hyperfine interaction and interaction with the external electric field. Finally, our Hamiltonian reads as
\begin{equation}
{\rm \hat{H}} = {\rm \hat{H}}_{\rm mol} + {\rm \hat{H}}_{\rm hfs} + {\rm \hat{H}}_{\rm ext},
\label{Hamtot}
\end{equation} 
where
\begin{equation}
 {\rm \bf\hat{H}}_{\rm hfs} = { g}_{\rm H} {\bf \rm I} \cdot \sum_a\left(\frac{\bm{\alpha}_a\times \bm{r}_a}{r_a^3}\right)
\end{equation}
is the hyperfine interaction between electrons and the hydrogen nucleus,
${ g}_{\rm H}$ is the
 g-factor of the hydrogen nucleus, $\bm{\alpha}_a$
 are the Dirac matrices for the $a$-th electron, $\bm{r}_a$ is its
 radius-vector in the coordinate system centered on the H nucleus,
index $a$ enumerates (as in all equations below) electrons of YbOH.

\begin{equation}
 {\rm \bf\hat{H}}_{\rm ext} =   -{ {\bf D}} \cdot {\bf E}
\end{equation}
describes the interaction of the molecule with the external electric field, and
{\bf D} is the dipole moment operator.

Wavefunctions were obtained by numerical diagonalization of the Hamiltonian (\ref{Hamtot})
over the basis set of the electronic-rotational-vibrational wavefunctions
\begin{equation}
 \Psi_{\Omega m\omega}P_{lm}(\theta)\chi(R)\Theta^{J}_{M_J,\omega}(\alpha,\beta)U^{\rm H}_{M_I}.
\label{basis}
\end{equation}
Here 
 $\Theta^{J}_{M_J,\omega}(\alpha,\beta)=\sqrt{(2J+1)/{4\pi}}D^{J}_{M_J,\omega}(\alpha,\beta,\gamma=0)$ is the rotational wavefunction, $U^{H}_{M_I}$ is the hydrogen nuclear spin wavefunctions and $M_J$ is the projection of the molecular (electronic-rotational-vibrational) angular momentum $\hat{\bf J}$ on the lab axis, 
 $\omega$ is the projection of the same momentum on $z$ axis of the molecular frame,
 $M_I=\pm1/2$ is the projection of the nuclear angular 
momentum $I=1/2$ of hydrogen on the same axis, $P_{lm}(\theta)$ is associate Legendre polynomial,
\begin{equation}
\Psi_{\Omega m\omega} = e^{i(\omega - 1/2\sum_a{\sigma_a})\varphi} \sum_i\Psi^e_{\Lambda_i}\Psi^s_{\Sigma_i},
\label{nexpansion}
\end{equation}
where $\sigma_a$ is electronic spin variables, $\Psi^e_{\Lambda_i}$ and $\Psi^s_{\Sigma_i}$ are orbital and spin wavefunctions respectively, 
$\Lambda_i$ and $\Sigma_i$ are projections of electronic orbital and spin momenta on $z$ axis, i enumerates all possible nonrelativistic wavefunctions.

Eq. (\ref{nexpansion}) is expansion of the electronic wavefunction in terms of nonrelativistic functions
of linear configuration. For linear configuration  only terms with $\Omega = \Lambda_i + \Sigma_i$ contribute. $\Psi_{\Omega m\omega}$ depends on $m$ since $\omega$ is chosen so that $\omega = \Omega +  m$. $\Psi_{\Omega m\omega}$ is eigenfunction of the electronic-vibrational momentum:
\begin{equation}
\hat{ J}^{e-v}_z \Psi_{\Omega m\omega} = \hbar \omega \Psi_{\Omega m\omega}.
\label{elvibeigen}
\end{equation}
To prove eq. (\ref{elvibeigen}), we note that nonrelativistic wavefunctions depend on 
$\varphi$ as $\Psi^e_{\Lambda_i} \sim e^{-i\Lambda_i\varphi}$ and
that $1/2\sum_a{\sigma_a}\Psi^s_{\Sigma_i} = \Sigma_i\Psi^s_{\Sigma_i}$.
For the lowest vibrational levels configuration of the molecule is close to linear, therefore the main contribution in sum (\ref{nexpansion}) is given by the terms with $\Omega = \Lambda_i + \Sigma_i$. Therefore, we also have approximate relations
\begin{equation}
\hat{ J}^{e}_z \Psi_{\Omega m\omega} \approx \hbar \Omega \Psi_{\Omega m\omega},
\label{elvibeigen2}
\end{equation}
\begin{equation}
\hat{ J}^{v}_z \Psi_{\Omega m\omega} \approx \hbar m \Psi_{\Omega m\omega}.
\label{elvibeigen3}
\end{equation}
In this  calculation functions with $\omega - m = \Omega = \pm 1/2$, $l=0-30$ (as in Ref. \cite{Zakharova:21b}) and $m=0,\pm 1, \pm 2$, $J=1/2,3/2,5/2$ were included to the basis set (\ref{basis}).
Note, that the ground mode $v=0$ corresponds to $m=0$,
the first excited bending mode $v=1$ to $m=\pm 1$, the second excited bending mode has states with $m=0, \pm2$ etc.
Using the basis set (\ref{basis}) allows us 
to employ (with a little modification) the codes developed for the diatomics calculation 
\cite{Petrov:11, Petrov:14, Petrov:15, Petrov:17, Petrov:18, Petrov:18b}, and
paves the way for computation of the g-factors, g-factors difference, different systematics in an external electro-magnetic field etc. These calculations are important for optimization of experimental conditions for the eEDM experiment.

In the present work we applied the method for calculation of sensitivity of $^{174}$YbOH to $\mathcal{P}$,$\mathcal{T}$-odd effects in the external electric field. As a test, we exactly reproduced with described method frequencies of bending and stretching vibrational modes (319 and 550 cm$^{-1}$ respectively)  and value 26 MHz of $l$-doubling for $^{174}$YbOH molecule obtained with the Hamiltonian (\ref{Hlabf}) in Ref. \cite{Zakharova:21b}.

For the purposes of the present paper, we found that ignoring the R dependence leads to an error of about 1\% or less for polarization $P$ of $^{174}$YbOH. Therefore, we reduced the Hamiltonian (\ref{Hmolf})
and basis set (\ref{basis}) to
\begin{equation}
\hat{\rm H}_{\rm mol}=\frac{(\hat{\bf J} -\hat{\bf J}^{e-v} )^2}{2\mu R^2}+\frac{\hat{\bf l}^2}{2\mu_{OH}r^2}+ {\rm  \hat{H}}_{\rm el},
\label{Hmolf2}
\end{equation}
and 
\begin{equation}
 \Psi_{\Omega m\omega}P_{lm}(\theta)\Theta^{J}_{M_J,\omega}(\alpha,\beta)U^{\rm H}_{M_I}
\label{basis2}
\end{equation}
with $R=3.9$ a.u. respectively.
In this approximation we neglect the influence of the stretching modes but nevertheless
take into account the bending ones with fixed $R$.
Provided that the {\it electronic-vibrational} matrix elements are known, the matrix elements of ${\rm \bf\hat{H}}$ between states in the basis set (\ref{basis2}) can be calculated with help of the angular momentum algebra \cite{LL77} in the same way as for the diatomic molecules \cite{Petrov:11}.
Note however, that pure electronic matrix elements, in general, depend on $\theta$,
and the selection rules for $\Omega$ quantum number can be violated. In turn the selection rules for $\omega$ quantum number are rigorous and the same as for $\Omega$ quantum number in the diatomics. For example the matrix element for the perpendicular hyperfine structure constant
\begin{multline}
\label{Aperp1}
A_{ \perp} = {g_{\rm H}} \times\\
   \langle
   \Psi_{\Omega=1/2m\omega}P_{lm} |\sum_a\left(\frac{\bm{\alpha}_a\times
\bm{r}_a}{r_a^3}\right)
_+|\Psi_{\Omega=1/2 m-1 \omega-1}P_{l'm-1}\rangle 
\end{multline}
is, in general, nonzero. Here
$$
\left(\frac{\bm{\alpha}_a\times \bm{r}_a}{r_a^3}\right)_+ = 
\left(\frac{\bm{\alpha}_a\times \bm{r}_a}{r_a^3}\right)_x + 
i\left(\frac{\bm{\alpha}_a\times \bm{r}_a}{r_a^3}\right)_y
$$
and similarly for other vectors.
Since hyperfine interaction on hydrogen is small, in the present work we neglected
their $\theta$ dependence and used the values
\begin{multline}
A_{ \parallel} = \frac{g_{\rm H}}{\Omega} \times\\
   \langle
   \Psi_{\Omega m\omega}P_{lm} |\sum_a\left(\frac{\bm{\alpha}_a\times
\bm{r}_a}{r_a^3}\right)
_z|\Psi_{\Omega m \omega}P_{l'm}\rangle \\
= 6.436~\delta_{ll'} {~ \rm MHz},
\end{multline}

\begin{multline}
A_{ \perp} = {g_{\rm H}} \times\\
   \langle
   \Psi_{\Omega=1/2m\omega}P_{lm} |\sum_i\left(\frac{\bm{\alpha}_i\times
\bm{r}_i}{r_i^3}\right)
_+|\Psi_{\Omega=-1/2 m \omega-1}P_{l'm}\rangle \\
= 3.977~ \delta_{ll'} {~ \rm MHz}
\end{multline}
from Ref. \cite{Pilgram:21}.
Spin-rotational interaction is modeled by matrix element \cite{Pilgram:21}
\begin{multline}
p=\frac{1}{\mu R^2}
\times\\
   \langle
   \Psi_{\Omega=1/2m\omega}P_{lm} |J^e_+
|\Psi_{\Omega=-1/2 m \omega-1}P_{l'm}\rangle \\
= 0.5116~ \delta_{ll'} {~ \rm cm}^{-1}.
\label{pme}
\end{multline}
The matrix element for the dipole moment operator
\begin{multline}
   \langle
   \Psi_{\Omega m\omega}P_{lm} | {\bf D}
|\Psi_{\Omega m \omega}P_{l'm}\rangle 
= 0.433~ \delta_{ll'} {~ \rm a.u.}
\label{dopvalue}
\end{multline}
was taken from Ref. \cite{prasannaa2019enhanced}.
Adiabatic potential was taken from Ref. \cite{Zakharova:21b}.

In an experiment to search the $\mathcal{P}$,$\mathcal{T}$-odd effects the opposite parity levels of the molecule are mixed in the external electric field to polarize the molecule. 
For a completely polarized ($P=\pm1$) $^{174}$YbOH the  $\mathcal{P}$,$\mathcal{T}$-odd interaction energy shift approach the maximum value
given by
$\delta E^{\rm max} = E_{\rm eff} d_e$.

\begin{figure}
\includegraphics[width=0.95\linewidth]{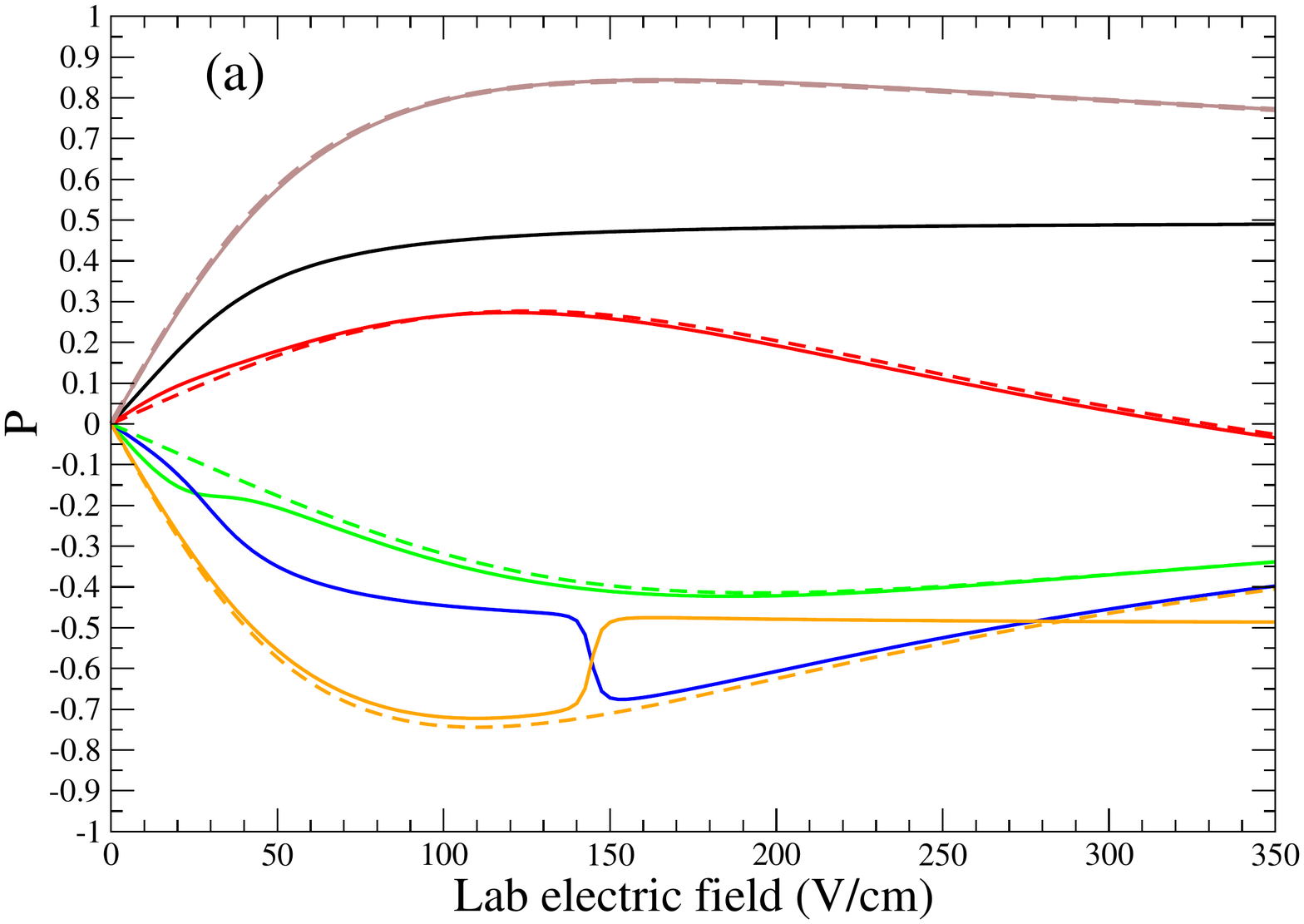}
\includegraphics[width=0.95\linewidth]{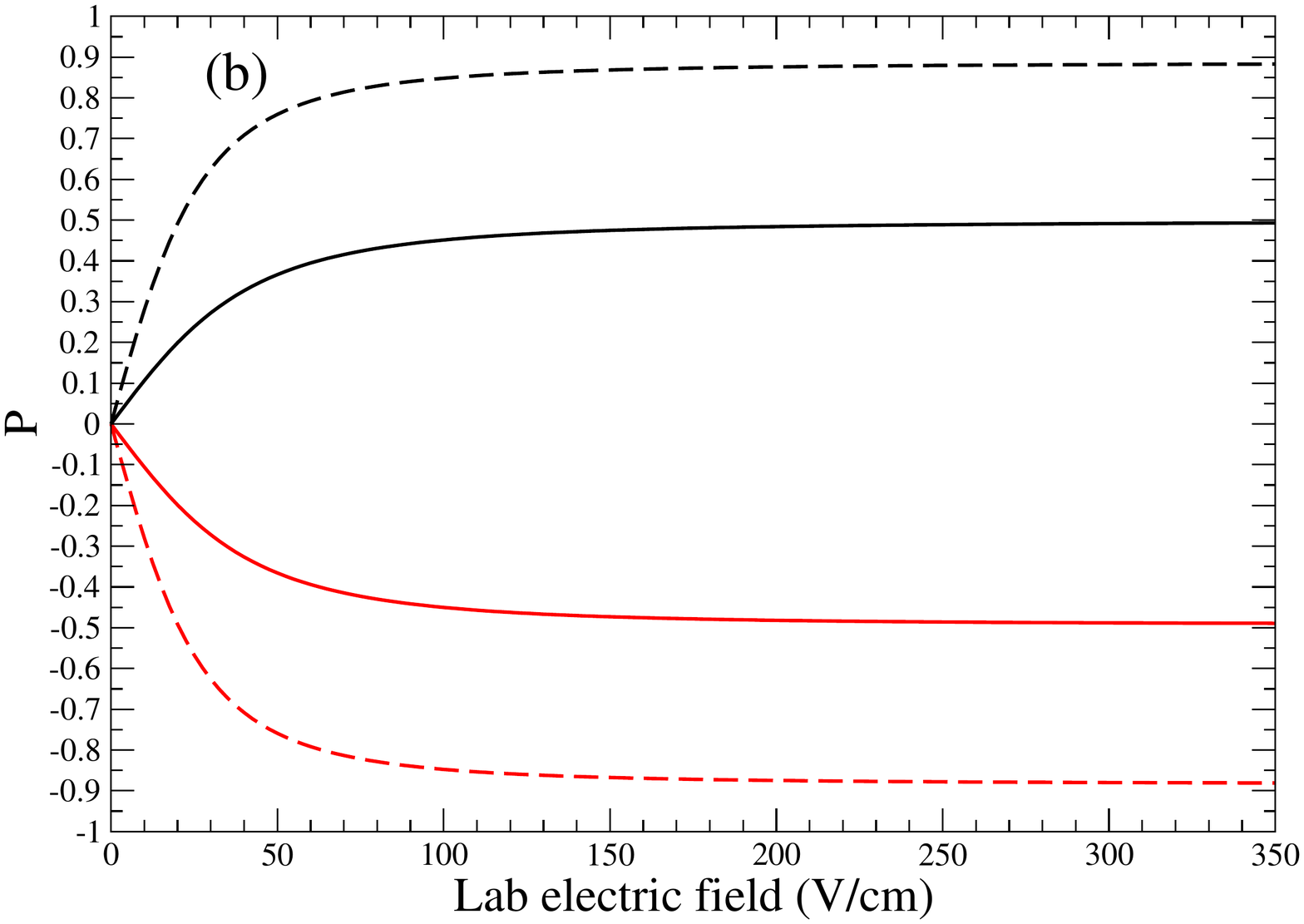}
\caption{\label{EDMshift} 
(Color online) Polarization $P$ (see eq. (\ref{split})) for the lowest $N=1$ rotational level
of the first excited the $v=1$ bending vibrational mode of $^{174}$YbOH as function of the external electric field
(a) Solid lines correspond to the $M_F=M_J+M_I=1$ levels.
Dashed lines correspond to the $M_J=1/2$ levels.
(b) $M_F=M+M_I=2$ levels. Solid lines correspond to the real matrix element (\ref{pme}),
dashed lines correspond to reduced matrix element (\ref{pme2})}
\end{figure}

Absolute value of the finite-electric-field shift (\ref{split}) is smaller than that of $\delta E^{max}$.
In Fig. \ref{EDMshift} the calculated polarization $P$ for the lowest $N=1$ rotational level
of the first excited $v=1$ bending vibrational mode of the $^{174}$YbOH as function of the external electric field is presented.
The function $E_{\rm eff}(R,\theta)$ from Ref. \cite{Zakharova:21b} was used for calculations.

Calculations showed that most of the levels have polarization equal to 50\% or less. In particular the stretched $M_F=M_J+M_I=2$ states smoothly approaches 50\% efficiency value. 
This is in contradiction with Ref. \cite{Kozyryev:17} where
about 100\% efficiency was found.

The reason for the fact that 
100\% efficiency is not reached is the large Coriolis interaction ($\frac{1}{2\mu R^2} J^e_+J_- $)
between levels $|\Omega=1/2,m=1,M_J=3/2,M_I=1/2,J=3/2\rangle$
and $|\Omega=-1/2,m=1,M_J=3/2,M_I=1/2,J=3/2\rangle$ (combination of above states gives $N=1$ state).
The interaction is of the same order of magnitude as the separation between close rotational levels. In the Hund's case (b) used in the Ref. \cite{Kozyryev:17} this fact can be explained by
expression $\langle mNSMM_S |\hat{n}| mNSMM_S\rangle = sgn(Mm)/N(N+1)$ for projection of the  vector $\hat{n}$ 
on laboratory axis. Here $\hat{\bf N} = \hat{\bf J} - \hat{\bf S}$, $\hat{\bf S}$ is electronic spin operator, $\hat{n}$ is the unit vector along the molecular axis,  $M,M_S=\pm1/2$ are the projections of the $\hat{\bf N}, \hat{\bf S}$ on the laboratory axis.
The factor $1/N(N+1) = 1/2$ for $N=1$ ($M=\pm 1$ for Hund's case ($b$)) gives 50\% suppression for the effect. The factor will be different for different rotational levels. For $N=2$, for example, it is 1/6 for $M=1$ and 1/3 for $M=2$. The same situation will be for other Hund's case $b$ molecules e.g. YbOCH$_3$, RaOH, etc. In Ref. \cite{Petrov:14} it is noted that $J= 2$ excited rotational state of ThO has the same sensitivity to $P,T$-odd effects as $J=1$, but is more robust against a number of systematic errors. Using excited rotational levels of YbOH one should take into account that sensitivity will be smaller than that for N=1 level.

The Spin-rotation interaction couples $\hat{\bf N}$ and $\hat{\bf S}$ to the $\hat {\bf J}$ angular momentum. The state $| m=1N=1SJ=3/2M_J\rangle = 3/4|\Omega=1/2,m=1,M_J,J=3/2\rangle
+ 1/4|\Omega=-1/2,m=1,M_J,J=3/2\rangle$ has $|P|=3/4-1/4 = 1/2$.
This is also seen from the consideration above and facts that
$| mN=1SJ=3/2M_J=3/2\rangle = | mN=1SM=1M_S=1/2\rangle $ and $P$ is independent of $M_J$. The state $| m=1N=1SJ=1/2M_J\rangle = |\Omega=-1/2,m=1,M_J,J=1/2\rangle$
 has $|P|= 1$. (Formulas for $m=-1$ can be obtained by substitution 
 $m \rightarrow -m, \Omega \rightarrow -\Omega$). The state $J=1/2$ with definite quantum number $m$ has maximum polarization $|P|=1$.
 However quantum number $m$ implies the mixing of $l-$doublets by some finite electric field.  Electric field also mixes $J=1/2$ and $J=3/2$ levels which leads to decreasing of the polarization
 up to $|P|=1/2$ value at the limit. Since at zero electric field $P=0$, the maximum of the $P$ value as function of electric field
 with  $1/2<|P|^{max}<1$ is observed. The larger is the spin-rotation interaction and smaller is the $l-$doubling the larger is $|P|^{max}$.

It is important to note that we considered projection of the  vector $\hat{n}$  on laboratory axis since in Hund's case $b$ quantization axis for the electron spin is an laboratory axis.
For Hund's case $a$ (when both orbital angular and spin angular momenta have definite projections on molecular axis) $^2\Pi_{1/2}$ states the matrix element of Coriolis interaction between $|\Omega=1/2,m=1,M_J=3/2,M_I=1/2,J=3/2\rangle$
and $|\Omega=-1/2,m=1,M_J=3/2,M_I=1/2,J=3/2\rangle$ is zero and $P=1$ value is reached. For Hund's case $c$ when $^2\Pi_{1/2}$ state has admixture of other $\Omega=1/2$ states (mostly $^2\Sigma$) Coriolis interaction is not zero again,
but less than one given by eq. (\ref{pme}).
In Fig. \ref{EDMshift} panel (b) the calculated polarization for matrix element
\begin{equation}
p = 0.15~ \delta_{ll'} {~ \rm cm}^{-1},
\label{pme2}
\end{equation}
corresponding to the structure of $X^2\Pi_{1/2}$ state of PbF \cite{Petrov:13} and can be expected on this level for PbOH molecule, is given.
As is expected, 
for a smaller value of $p$, larger polarization $P$ is reached.

For diatomic molecules, for a given value of external electric field, a smaller value of $p$ implies larger $P$ value as well. We stress however, that for triatomics we discuss here the saturated value of $P$ which is independent of $p$ in diatomics.

 Exact values  which takes into account mixing of the levels by the hyperfine interaction should be obtained in molecular calculations as described in our method.
Solid lines on Figure \ref{EDMshift} panel (a) show that four of the six levels
$M_F=1$ similarly to $M_F=2$ have polarization less than 50\%. Two of them ($m=\pm 1, J=1/2$ levels) reach values 84\% and -72\% for electric fields 109 and 166 V/cm respectively and their absolute values decrease for higher values of the field. Dashed lines on Figure \ref{EDMshift} panel (a) correspond to
 $M_J=1/2$ states assuming zero nuclear spin (and therefore hyperfine interaction is neglected).
 One sees that hyperfine interaction only little change the $P$ value.
 
\begin{figure}
\includegraphics[width=0.95\linewidth]{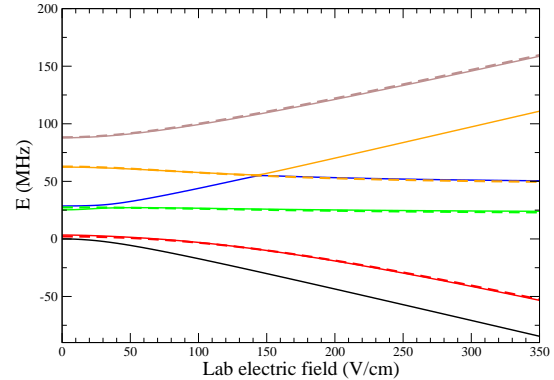}
\caption{\label{Ener} 
(Color online) Energies of the lowest $N=1$ rotational level
of the first excited $v=1$ bending vibrational mode of $^{174}$YbOH as functions of the external electric field.
Solid lines correspond to the $M_F=M_J+M_I=1$ levels.
Dashed lines correspond to the $M_J=1/2$ levels.
Colors of lines correspond to colors of lines in Fig. \ref{EDMshift} panel (a).}
\end{figure}

For dipole moment (\ref{dopvalue}) increased (decreased) on 10\% the extreme values for polarization remain the same but corresponding values for electric field shifted to 100 and 150 V/cm (122 and 184 V/cm).  
Energy levels of the six levels $M_F=1$ and four level $M_J=3/2$ (assuming zero nuclear spin) on Figure \ref{Ener} are shown.
At zero electric field the first four $M_F=1$ levels are $J=3/2$ ones  and the next two are $J=1/2$ levels.
At electric field strength of about 150 V/cm the avoided crossing between fourth and fifth levels $M_F=1$ is observed. Due to the large changing of the corresponding wave functions sharp change in $P$ of these levels is observed.

Finally we considered the polarization of $^{174}$YbOH molecule in the first excited bending mode by the external electric field. We found that values for polarization are smaller than 100\% for realistic electric field strengths and decreases as rotational quantum number increases. The limit for polarization is  $P=Mm/N(N+1)$ (e.g. $P=1/2$ for N=1, 
$P=1/6,1/3$ for N=2 etc). The spin-rotation interaction increase 
$P$ value forming a maximum at some electric field.
The same situation will be for other Hund's case $b$ molecules e.g. YbOCH$_3$, RaOH, etc. Hyperfine interaction can potentially increase $P$ value up to 100\%.
For $^{174}$YbOH $P=0.84$ is reached for electric field strength 109 V/cm.
Results of our letter should be used for correct extracting of eEDM value from YbOH experiment and optimal choice of the quantum state and electric field strength. Our results confirm fundamental importance of polyatomic molecules as candidates for eEDM measurements.

The authors thank Nicholas Hutzler, Lasner, Zack and Arian Jadbabaie for useful discussion.
The work is supported by the Russian Science Foundation grant No. 18-12-00227.

%
%

\end{document}